\pdfoutput=1
\documentclass[aps,prl,reprint,superscriptaddress,nofootinbib,longbibliography]{revtex4-2}

\usepackage{amsmath,amssymb}
\usepackage{graphicx}
\usepackage{xcolor}
\usepackage[normalem]{ulem} %
\usepackage{hyperref}
\graphicspath{{letter/figures/}}

\definecolor{pur}{rgb}{0.6,0.,1.}

\begin{document}

\title{Forward-Backward Asymmetries from Bottom Flux Correlators\\ with DELPHI Open Data}

\author{Sanghyun Ko}
\affiliation{CERN, 1211 Geneva 23, Switzerland}

\author{Kyle Lee}
\affiliation{High Energy Physics Division, Argonne National Laboratory, Lemont, Illinois 60439, USA}

\author{Dietrich Liko}
\affiliation{Marietta Blau Institute for Particle Physics, Austrian Academy of Sciences, 1010 Vienna, Austria}

\author{Ian Moult}
\affiliation{Department of Physics, Yale University, New Haven, Connecticut 06511, USA}

\author{Sitian Qian}
\affiliation{Department of Physics and Astronomy, Northwestern University, Evanston, Illinois 60208, USA}

\author{Jingyu Zhang}
\affiliation{Department of Physics and Astronomy, Vanderbilt University, Nashville, Tennessee 37235, USA}

\date{\today}

\begin{abstract}
Measurements of forward-backward asymmetries at electron-positron colliders have provided some of the most precise measurements of the electroweak sector, and constraints on beyond the Standard Model physics. While global electroweak fits are largely consistent, a longstanding tension has persisted in the forward-backward asymmetry of heavy quarks, motivating new theoretical approaches and experimental measurements. Recently, a new theoretical approach to forward-backward asymmetries has been proposed, formulating them as one-point correlators of conserved fluxes. In this \emph{Letter}, we perform the first measurement of the one-point function of bottom flux, using archival data from the DELPHI experiment. We fit the observed asymmetry and compare its coefficient with perturbative calculations to extract the effective weak mixing angle, $\sin^2\theta_{\mathrm{eff}}^{\ell}$. This establishes the experimental feasibility of this approach, and opens the door to a re-invigorated electroweak precision program with archival data.
\end{abstract}

\maketitle

\section{Introduction}

The symmetry breaking pattern of the Standard Model famously leads to unambiguous predictions, computed to high theoretical accuracy, for relations between experimental observables. With the discovery and direct measurement of the mass of the Higgs, global precision electroweak fits provide overconstrained fits and offer some of our strongest constraints on physics beyond the Standard Model.

While broadly consistent, observables related to forward-backward asymmetries of heavy quarks exhibit a longstanding tension \cite{Baak:2014ora,Haller:2018nnx,Fischer:2026bka}. On the one hand, modifications in observables related to the third generation are particularly natural in many models of physics beyond the Standard Model \cite{Haber:1999zh,Peccei:1990uv,Choudhury:2001hs,Morrissey:2003sc,Kumar:2010vx,Batell:2012ca,He:2002ha,Djouadi:2006rk,DaRold:2010as,Liu:2017xmc,Glioti:2024hye,Allwicher:2023shc,Davighi:2023iks}, leading to optimism that this might be a first hint of new physics. On the other hand, unlike leptonic asymmetries, heavy quark asymmetries are dominated by the physics of quantum chromodynamics (QCD), making it plausible that uncertainties on theoretical predictions are underestimated. Either way, new experimental and theoretical studies of forward-backward asymmetry observables with independent uncertainties are highly motivated to resolve this longstanding issue.

\begin{figure}
\centering
\includegraphics[width=0.4\textwidth]{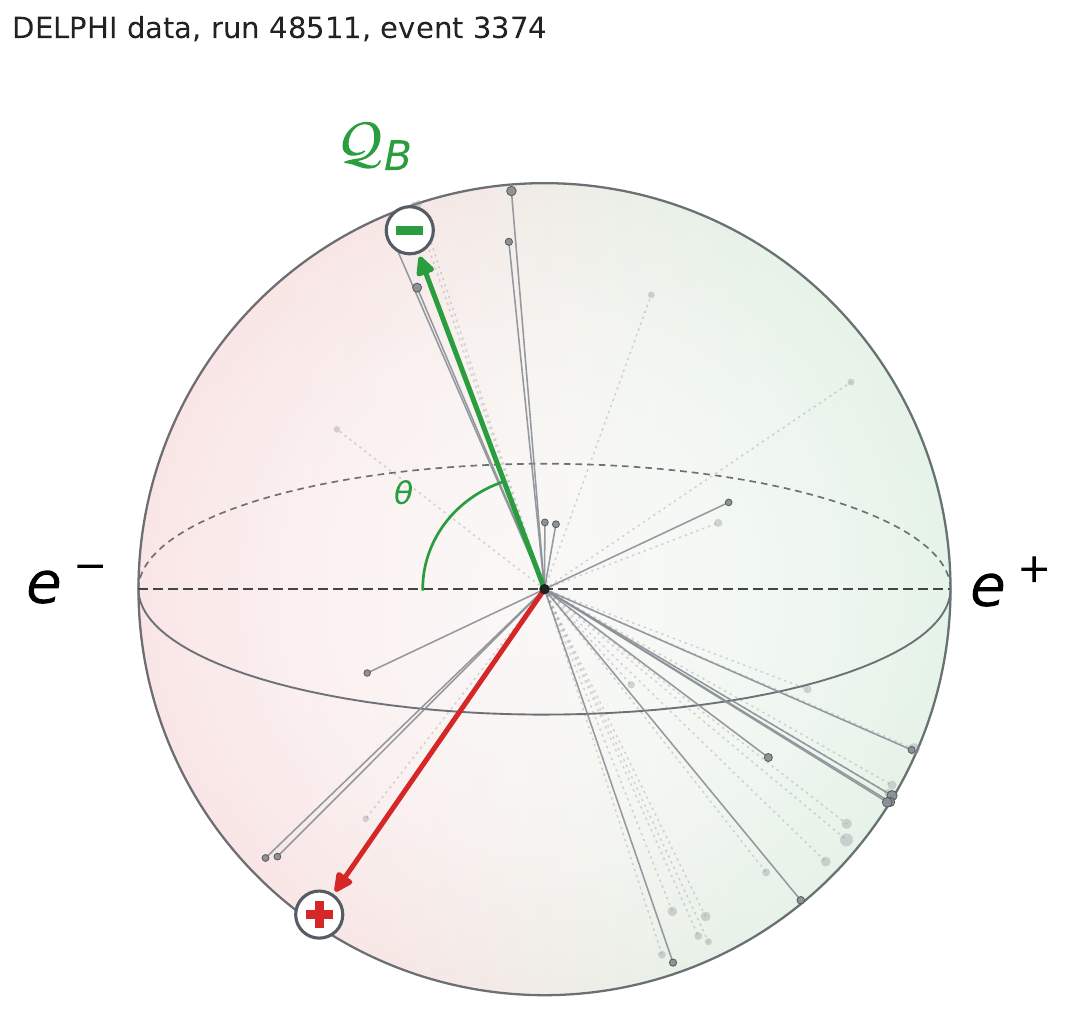}
\caption{An illustration of the measurement of heavy flavor flux in electron-positron collisions. Due to a mixed electroweak anomaly in the Standard Model, flavor flux exhibits a characteristic forward-backward asymmetry with respect to the beam.}
\label{fig:detector_image}
\end{figure}

A recent theoretical proposal \cite{Lee:2026bcq}, building on the seminal work of \cite{Hofman:2008ar}, showed that forward-backward asymmetries of heavy quarks can be elegantly formulated in terms of one-point correlators of conserved flavor charges. Remarkably, this directly relates them to mixed electroweak anomalies in the Standard Model, providing a clean field theory interpretation, and enabling precision calculations. From a practical perspective, this approach also overcomes a number of the ambiguities inherent in more traditional definitions of quark forward-backward asymmetries. Previous measurements \cite{OPAL:1993wua,ALEPH:2001mdb,L3:1992fsb,DELPHI:2004wzo} required the reconstruction of an (inherently ambiguous) quark direction and charge, which was typically defined using jet definitions or hemispheres \cite{OPAL:1993wua,L3:1992fsb,ALEPH:2001mdb,DELPHI:2003fml,DELPHI:2004wzo,OPAL:2003pfe,DELPHI:1991mqi,OPAL:1992jsm,ALEPH:1996qlh,ALEPH:1998pmr}. The use of jet or hemisphere charge introduces sensitivity to non-perturbative corrections. For a discussion of QCD uncertainties, see \cite{dEnterria:2018jsx,AlcarazMaestre:2020fmp}. The approach of \cite{Lee:2026bcq} promises to overcome these issues, however, correlators of bottom flux have never been measured in experiment.

In this \emph{Letter}, we use archival DELPHI data from the LEP collider to perform the first measurement of the one-point correlator of bottom flux. We develop a simple b-tagging procedure, which we deploy on the archival data, and achieve a measurement precision of $\sim 10\%$. We fit the observed asymmetry and compare its coefficient with perturbative calculations to extract the effective weak mixing angle, $\sin^2\theta_{\mathrm{eff}}^{\ell}$. Our results demonstrate the experimental feasibility of this approach and illustrate the great potential of archival data for precision electroweak measurements.

\section{Forward-Backward Asymmetries from Flavor Flux Correlators}
\label{sec:observable}

Forward-backward asymmetries have an illustrious history dating back to the earliest lepton colliders. Leptonic forward-backward asymmetries are extremely clean experimental observables since the outgoing leptons can be directly measured in asymptotic detectors. Quark forward-backward asymmetries on the other hand are more challenging to formulate since quarks do not appear as asymptotic states in detectors. Instead, one must design observables which capture the angular distribution of the $Z\to q\bar q$ vertex from the fluxes of hadrons in the detector. Ideally, any such proposed observable should be theoretically calculable, and directly measurable in experiment.

There has been significant recent progress in the understanding of asymptotic observables \cite{Moult:2025nhu}. This approach was first presented in the work of \cite{Hofman:2008ar}, which formulated the measurement of asymptotic fluxes in terms of correlation functions of currents. One example where this has proven particularly successful is the study of energy flux, for example the precise measurement of energy correlators using archival LEP data \cite{Electron-PositronAlliance:2025fhk,Jaarsma:2025tck}. In the context of conformal field theories, \cite{Hofman:2008ar} made the particularly intriguing observation that forward-backward asymmetries are related to anomalous three-point functions. This was generalized to the case of charge flux in the Standard Model in \cite{Zhang:2026emt}, and shown to be measureable at electron-positron colliders \cite{Zhang:2026vdo,Zhang:2026esd}, highlighting a macroscopic manifestation of quantum field theoretic anomalies.

In  \cite{Lee:2026bcq}, a generalization to flavor fluxes was presented, providing an opportunity to formulate heavy quark forward-backward asymmetries in terms of correlation functions and anomalies. A $b$-flavor detector can be defined as
\begin{align}
\label{eq:detector}
\mathcal{Q}_b(\vec{n}) = \lim_{r\to\infty} r^2\int_0^\infty d t\; n_i\, J_b^i(t, r\vec{n})\,.
\end{align}
Experimentally, this is implemented by the action of the $b$-flavor detector on hadronic states
\begin{align}
\label{eq:detector_action}
\mathcal{Q}_b(\vec{n})\,|X\rangle = \sum_{h\in X} b(h)\, \delta^{2}(\vec{n}-\vec{n}_h)\,|X\rangle\,,
\end{align}
where $b(h)$ returns the $b$-quantum number of the hadron observed in the detector.

In the case of measurements at the $Z$-pole, this reformulation is particularly elegant. At the $Z$-pole, the colliding electron and positron couple to the QCD vacuum via the $Z$-boson current. The one-point correlator of the $b$-flux is therefore related to a three-point function of currents $\langle J_Z J_b J_Z \rangle$, a particularly well studied theoretical object. Indeed, in the case of a conformal theory, such three-point functions are fixed by an anomaly coefficient \cite{Schreier:1971um,Osborn:1993cr}, leading to the form
\begin{align}
\langle \mathcal{Q}(\vec{n})\rangle_{J}&
= \frac{3}{\pi}\frac{d_{J\! J Q}}{c_{J}}\frac{i\epsilon^{ijk} \epsilon^*_i \epsilon_j n_k}{\epsilon^*_l \epsilon_l}\,, %
\label{eq:oneptCharge}
\end{align}
for any charge correlator \cite{Hofman:2008ar}. Here $\epsilon$ is the polarization vector of the current, which in the lab frame corresponds to the $\cos(\theta)$ angular asymmetry of the forward-backward asymmetry. This provides a beautiful rephrasing of heavy quark forward-backward asymmetries at the $Z$-pole as macroscopic manifestations of anomalies. This was extended to the case of non-conformal theories in \cite{Zhang:2026emt}, where it was shown that the characteristic angular form of Eq. \ref{eq:oneptCharge} holds, with normalization corrections proportional to the $\beta$ function. It was shown in \cite{Lee:2026bcq} that the hadronic bottom-flux detector in Eq.~(\ref{eq:detector_action}) can be matched onto the partonic bottom-flux detector in perturbation theory because bottom charge is gapped at the scale $m_b\gg\Lambda_{\mathrm{QCD}}$. This established the one-point $b$-flux correlator as a promising reformulation of forward-backward asymmetries.

\section{Experimental Analysis with DELPHI Archival Data}
\label{sec:delphi}
By Eq.~(\ref{eq:detector_action}), the bottom flux counts every hadron at its own direction, weighted by $b(h)$, which is $+1$ ($-1$) for a hadron that contains a $b$ ($\bar b$) quark, opposite in sign to the PDG bottomness~\cite{ParticleDataGroup:2026mpi}. Its one-point correlator is therefore a signed cross section, the angular distribution of the hadrons with $b(h)=+1$ minus that of the hadrons with $b(h)=-1$, and measuring it amounts to counting $b$ hadrons with their directions and their signs. We use the $e^+e^-$ collision data recorded by the DELPHI detector~\cite{DELPHI:1990cdc,DELPHI:1995dsm} at LEP in 1994 at $\sqrt{s}=91.2$~GeV. The data and DELPHI's simulation are public on the CERN Open Data portal~\cite{DELPHI:OD:short94c2,DELPHI:OD:kk2fqqpy_l94_c2}, under the collaboration's open-access policy~\cite{DELPHI:OD:policy}. The parts of the detector that matter here are the silicon vertex detector with three layers, the time projection chamber in a solenoidal field of 1.23~T, and the identification of electrons and muons. The polar angle $\theta$ is measured from the direction of the $e^-$ beam. We also use samples of $Z\to b\bar b$, $c\bar c$ and light-quark events generated with PYTHIA~8~\cite{Bierlich:2022pfr} and passed through DELPHI's detector simulation and reconstruction~\cite{SimulationWorkingGroup:1988sdz,SimulationWorkingGroup:1988eya,DELPHI:simana94c}. The simulation serves only to subtract the lighter flavors, to perform a closure test of the calibration, and to estimate the residual effects discussed below.

DELPHI reprocessed its data into a final short-DST format that includes all the corrections known at the time, in the IDEA project that was set up to ensure long-term access to them~\cite{Camporesi:2000xqf}. We analyze the 1994 data in this form, read with DELPHI's own software and converted to EDM4hep without reprocessing~\cite{Gaede:2022leb,delphi-edm4hep:2026}. They show a shift of the track momentum scale, which DELPHI itself observed as a shift of the $D^0$ mass~\cite{DELPHI:1998oyl}. We correct it with $K^0_S\to\pi^+\pi^-$ decays and $Z\to\mu^+\mu^-$ events, with a parametrization that includes a charge-odd term, as described in the companion paper. Hadronic $Z$ decays are selected with the standard DELPHI criteria~\cite{DELPHI:2003yqh}, based on charged particles with a momentum above 0.4~GeV and a polar angle between $20^\circ$ and $160^\circ$. We further require at least seven charged particles, a summed energy of the accepted particles above 15~GeV, and a reconstructed primary vertex. The selection keeps 1\,354\,997 events.

The flux requires, for each $b$ hadron, its direction and the sign of its bottom quantum number at production, $b(h)=+1$ for a hadron that contains a $b$ quark. Both are estimated from the decay vertex of the hadron and its jet. The candidates are the secondary vertices found by DELPHI's $b$-tagging package~\cite{DELPHI:2002tzi}, which clusters the particles of an event into jets with the JADE algorithm at $y_{\rm cut}=0.01$~\cite{JADE:1986kta,JADE:1988xlj} and searches each jet for one secondary vertex. The jets are used to group the tracks, and no jet axis is taken as the direction of a hadron. We keep the vertices with at least three tracks, a distance from the primary vertex above six times its uncertainty, and an invariant mass of the tracks above 1.5~GeV, which suppresses charm. The summed momentum of the tracks associated with the vertex gives the direction. In the simulation, its polar angle is reconstructed without bias and with a resolution of $1.2^\circ$. The data contain 191\,826 candidates: 128\,614 events have one candidate (single-tagged) and 31\,606 have two (double-tagged). As estimated from the simulation, 91\% of the candidates are $b$ hadrons, 7\% are charm hadrons, and 2\% come from light-quark events. The sign is taken from one of two decay products. If the jet of the vertex contains a track identified as an electron or a muon, with a transverse momentum above 1~GeV with respect to the jet axis, $b(h)$ is estimated as minus the lepton charge, as in the decay $b\to\ell^-$. Otherwise, it is minus the sign of the summed charge of the vertex tracks, when this is not zero, as for a charged $B$ meson. The other candidates are unsigned. Either sign can be wrong, because of cascade decays $b\to c\to\ell^+$, the oscillation of the neutral $B$ mesons, and lost or wrongly associated tracks. These effects are corrected together by the in-situ calibration.

The signed count of the candidates in each bin of $\theta$ measures the flux reduced by two properties of the tagger: the efficiency with which a $b$ hadron is found and signed, and the dilution factor $D=1-2w$, where $w$ is the probability of a wrong sign. Every signed candidate enters this count, from the single-tagged and the double-tagged events alike, and the double-tagged events serve in addition to calibrate the tagger. The flux is obtained by subtracting the contribution of the lighter flavors, dividing the signed count by the efficiency and the dilution factor, and normalizing to the number of $b\bar b$ events, taken as the number of selected events times $R_b$~\cite{ParticleDataGroup:2026mpi}. The efficiency and the dilution factor are both measured on the data, for each of the two sign estimators separately. The lepton signs 3\% of the $b$ hadrons, with a dilution factor of 0.50, and the vertex charge 18\%, with a dilution factor of 0.37. The efficiency follows from a tag-and-probe, that is, from how often an event with one tagged hadron is double-tagged. For the dilution factor, equal signs of the two candidates of a double-tagged event mean that one of them is wrong, so the excess of opposite-sign over same-sign pairs, as a fraction of all pairs with two signs, is the product of the dilution factors of the two candidates. An oscillation produces a same-sign pair just as a wrong sign does, so the dilution factor includes the mixing and no separate correction is applied. Both quantities are measured in twelve bins of $\theta$ between $30^\circ$ and $150^\circ$. In a closure test on simulated $Z\to b\bar b$ events, the procedure returns $1.021\pm0.048$ times the true flux. The derivation of the calibration and its tests are given in the companion paper.

The measurement and the calibration use the same events, so the statistical uncertainties are obtained from 2000 bootstrap resamples of the whole procedure. For the amplitude of the flux, the statistical uncertainty is 9.0\%, and that of the calibration is 5.5\%. The effects that the calibration does not measure are estimated with the PYTHIA~8 simulation. They are the correlations between the two tagged hadrons of an event, in their efficiencies and in their sign errors, the preference of the selection for energetic $b$ hadrons, and the migration between the bins of $\theta$. Each changes the amplitude by less than 2.5\%, and together they change it by $-0.3\%$. A charge-dependent difference between the forward and backward parts of the detector, which would imitate a flux, is controlled by the charge-odd term of the momentum calibration. The simulation enters the result through the subtraction of the lighter flavors. Replacing PYTHIA~8 by DELPHI's own simulation~\cite{DELPHI:OD:kk2fqqpy_l94_c2} changes the amplitude by 1.6\%. The change is consistent with zero within the statistics of the simulated samples and is taken in full as the systematic uncertainty. The total uncertainty is 10.6\%. The measurement is limited by its statistics, and the details are given in the companion paper.

\section{Results}
\label{sec:results}
\begin{figure}
\centering
\includegraphics[width=\columnwidth]{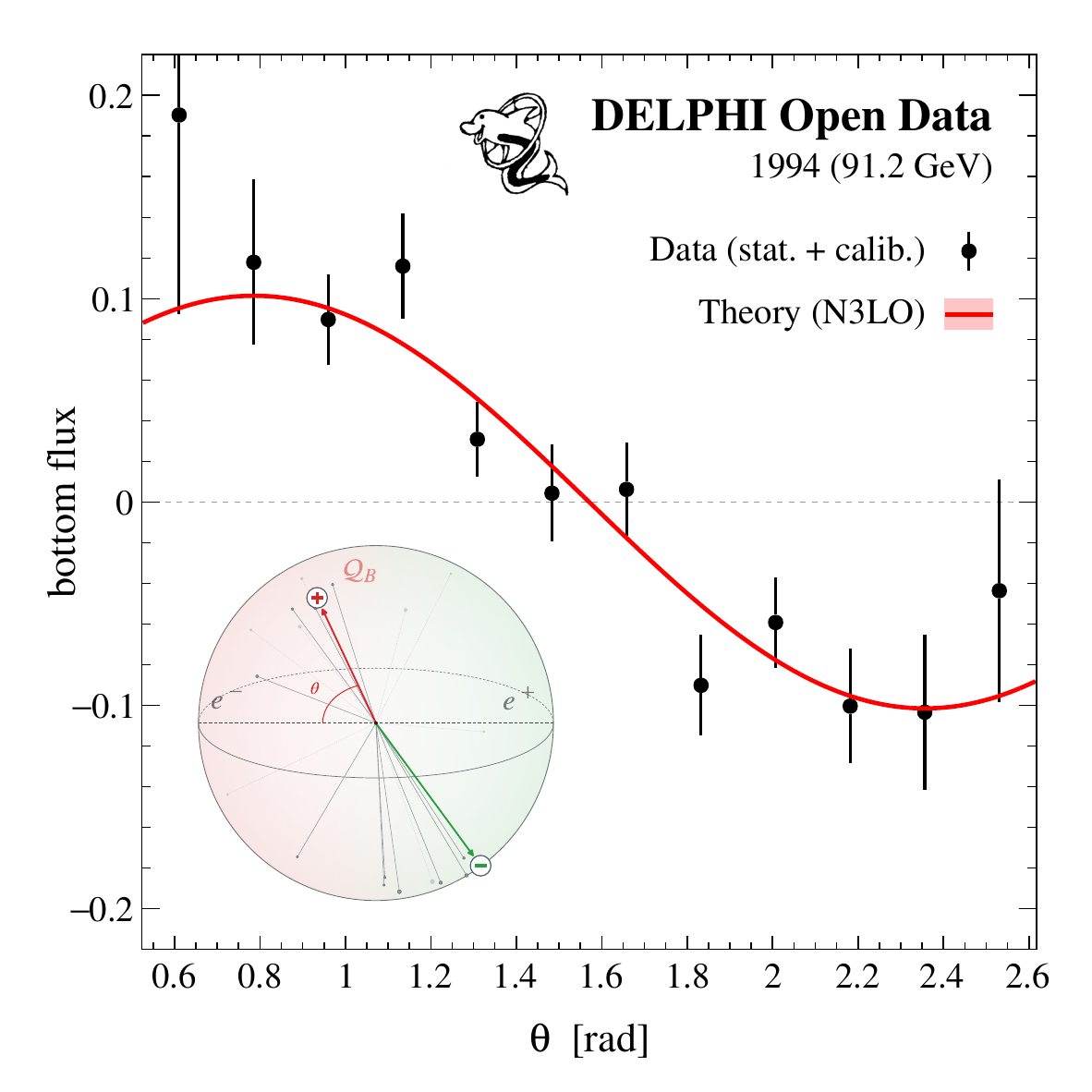}
\caption{Measured bottom-flux distribution using archival 1994 DELPHI data at $\sqrt{s}=91.2$~GeV. The black points show the data with statistical and calibration uncertainties. The red curve is the theoretical angular form evaluated at the extracted $\sin^2\theta_{\mathrm{eff}}^{\ell}=0.23006$, with $\alpha_s(m_Z)=0.1180$. Its shaded band shows the perturbative scale variation at fixed extracted angle described in the text.}
\label{fig:results}
\end{figure}

After integrating over the azimuthal angle and normalizing to the number of $b\bar b$ events, the theoretical angular dependence of the bottom flux is
\begin{align}
\frac{2\pi\sin\theta}{R_b}\langle\mathcal{Q}_b(\vec n)\rangle
= C_b\cos\theta\sin\theta\,.
\label{eq:flux_angular_form}
\end{align}
The factor of $\sin\theta$ is the Jacobian from $\cos\theta$ to $\theta$. The measured distribution in Fig.~\ref{fig:results} is consistent with this form. Its fitted coefficient, twice the amplitude of the $\sin2\theta$ modulation of the flux, is
\begin{align}
\begin{aligned}
C_b^{\mathrm{exp}}&=0.2029\pm0.0182\;\text{(stat.)}\\
&\quad\pm0.0112\;\text{(calib.)}\pm0.0032\;\text{(syst.)}\,.
\end{aligned}
\label{eq:flux_coefficient}
\end{align}
The first uncertainty is statistical, the second is that of the in-situ calibration of the tagger, and the third is systematic, from the subtraction of the lighter flavors. Combined in quadrature, they amount to $\pm0.0216$.
Before comparison with theory, we estimate the effect of QED radiation, dominated by initial-state radiation, using the quark-asymmetry correction quoted in Ref.~\cite{Freitas:2004mn}. We then obtain the radiation-corrected measurement coefficient $C_b^{\mathrm{corr}}\simeq0.2104\pm0.0216$. We compare this result with theory at the collision energy $Q=91.2$~GeV.

We compare this corrected coefficient with the perturbative bottom-flux calculation of Ref.~\cite{Lee:2026bcq} to extract the weak mixing angle. We fix $\alpha_s(m_Z)=0.1180$ and parametrize the electron and bottom vector couplings as $v_e=-\tfrac12+2x$ and $v_b=-\tfrac12+\tfrac23x$, where $x\equiv\sin^2\theta_{\mathrm{eff}}^{\ell}$.
The calculation is evaluated at $Q=91.2$~GeV to match
the collision energy of the data, with $m_Z=91.1876$~GeV
and $\Gamma_Z=2.4952$~GeV in the $Z$ propagator.
Photon exchange and $\gamma$--$Z$ interference are retained
in the hard-scattering coefficients.
Solving
$C_b^{\mathrm{th}}(Q,x,\alpha_s)=C_b^{\mathrm{corr}}$
for $x\equiv\sin^2\theta_{\mathrm{eff}}^{\ell}$, using the
perturbative prediction from Ref.~\cite{Lee:2026bcq}, gives
\begin{align}
\sin^2\theta_{\mathrm{eff}}^{\ell}
&=0.23006\pm0.00205\;\text{(exp.)}\nonumber\\
&\quad{}^{+0.000047}_{-0.000046}\;\text{(th.)}\,.
\label{eq:weak_angle_extraction}
\end{align}
The experimental uncertainty combines the statistical, calibration, and systematic contributions in quadrature and propagates them through the extraction. The theory uncertainty shown here is the QCD renormalization-scale variation described below; it does not include uncertainty in the adopted radiation correction or its transfer to the flux observable.

\begin{figure}
\centering
\includegraphics[width=\columnwidth]{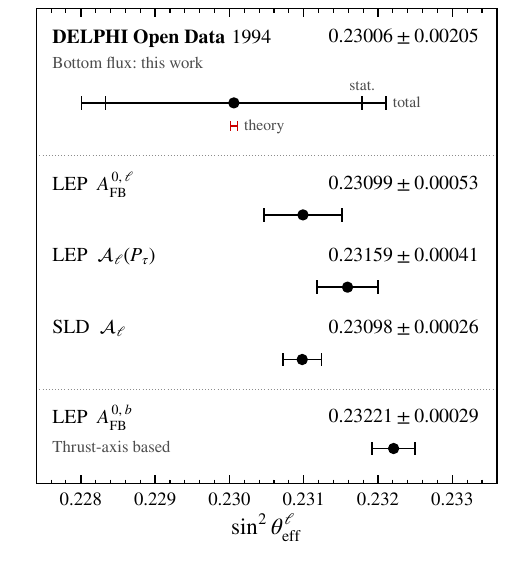}
\caption{
The effective weak mixing angle extracted from the DELPHI bottom-flux measurement, compared with published determinations~\cite{ALEPH:2005ab}. The middle rows are from leptonic observables at LEP and SLD. The value from the $b$-quark forward-backward asymmetry at LEP is also shown, because it is measured with respect to the thrust axis and relies on corrections for QCD radiation and hadronization. For the DELPHI point, the inner ticks mark the statistical uncertainty and the outer ones the total, with the statistical, calibration, and systematic uncertainties combined in quadrature. The red interval below the DELPHI point shows the perturbative scale variation described in the text. The other intervals are the published total uncertainties.}
\label{fig:weak_angle}
\end{figure}

Figure~\ref{fig:results} also shows the N$^3$LO theoretical flux evaluated at the extracted angle, with an estimated QED radiation correction~\cite{Freitas:2004mn} applied to its normalization for comparison with the measured distribution. Its central normalization therefore agrees with the measured coefficient by construction. The shaded band varies the renormalization scale over $\mu_R/Q=\{1/2,1,2\}$ at fixed extracted angle, including the explicit renormalization logarithms. This variation is much smaller than the experimental uncertainty in the flux. The effect of hadronization on the amplitude of the flux, evaluated with simulation, is negligible. Repeating the weak-angle extraction at these scales gives the red interval in Fig.~\ref{fig:weak_angle}, corresponding to shifts of about $\pm4.6\times10^{-5}$. These bands estimate the perturbative scale uncertainty; they have no assigned confidence level and do not include all power corrections or electroweak matching effects.

In Fig.~\ref{fig:weak_angle}, we compare our result with published determinations from leptonic observables~\cite{ALEPH:2005ab}. They are $0.23099\pm0.00053$ from the forward-backward asymmetry and $0.23159\pm0.00041$ from the tau polarization at LEP, and $0.23098\pm0.00026$ from the left-right asymmetry at SLD. We also show the LEP determination from the $b$-quark forward-backward asymmetry, $0.23221\pm0.00029$, set apart in the figure. As discussed in the Introduction, it is measured with respect to the thrust axis and requires corrections for QCD radiation and hadronization, making its relation to the weak mixing angle more dependent on non-perturbative modeling~\cite{dEnterria:2018jsx,AlcarazMaestre:2020fmp}. Our extraction is consistent with all of these determinations, although their measurement uncertainties are substantially smaller. Our uncertainty is dominated by statistics, including the statistics of the in-situ calibration. The measurement uses one year of data from one experiment, less than a tenth of the hadronic $Z$ decays recorded at LEP, and a simple tagger. The full LEP data set would reduce the uncertainty by a factor of about three, and a more powerful tagger would reduce it further.

\section{Conclusions}
\label{sec:conclusions}

In the past several years there has been tremendous progress rephrasing questions in collider physics in terms of precise field theoretic objects, namely correlations in fluxes of charge and energy. Many of these observables have now been measured in experiment. In this \emph{Letter}, we have achieved this synthesis for forward-backward asymmetries of heavy quarks. We showed, using archival DELPHI data, that the formulation of forward-backward asymmetries as a one-point correlator of bottom flux \cite{Lee:2026bcq}, can be directly measured in electron-positron experiments, and performed the first measurement of this observable. This opens the door to precision electroweak studies using archival data combined with new theory techniques. Beyond LEP, our approach can also be performed using archival SLD data. It also serves as an important testing ground for future lepton colliders. Combined, we are optimistic about the potential for resolving longstanding tensions in the electroweak sector of the Standard Model.

\emph{Acknowledgments.}---We benefited from discussions with members of the DELPHI Collaboration involved in its data preservation and are especially grateful to Ulrich Schwickerath, whose sustained efforts to preserve the data and software and make them publicly available and usable made this work possible. We thank Michael Peskin, Huilin Qu, Manqi Ruan, and Ben Nachman for useful discussions.
K.L. is supported by the U.S. Department of Energy under contract DE-AC02-06CH11357.
I.M. is supported by the DOE Early Career Award DE-SC0025581, the Sloan Foundation, and the Simons Collaboration on Confinement and QCD Strings.
S.Q. is supported in part by the U.S. CMS Operations Program.
J.Z. is supported in part by the U.S. National Science Foundation under ID 2514008.

\emph{AI Use Disclosure.}---The authors used the AI assistants Claude (Anthropic; versions Opus~5.5, Fable~5.1, and Fable~5.2) and Codex (OpenAI; versions 6~Astra and 5.5~Sol) in this work. Under the authors' direction, they assisted in writing the analysis software and the plotting scripts, in drafting and editing parts of the text, and in checking the manuscripts for consistency. The authors defined the method, reviewed the results, and revised all text, and they take full responsibility for the content.

\bibliography{PRL_version2}

\end{document}